\documentclass[aps,floats,epsf,twocolumn]{revtex4}
\usepackage{graphics}

\begin{document}

\title{Finite temperature superfluid density in very underdoped cuprates}

\author{Igor F. Herbut and Matthew J. Case}

\affiliation{Department of Physics, Simon Fraser University, 
Burnaby, British Columbia, Canada V5A 1S6 }

\begin{abstract}
The combination of a large gap, low transition temperature, and quasi 
two-dimensionality in strongly underdoped high temperature
superconductors severely constrains
the behavior of the $ab$-plane superfluid density $\rho$ with temperature $T$.
In particular, we
argue that the contribution of nodal quasiparticles to $\rho(T)$
is essential to account both for the amplitude of, and
the recently observed deviations from, the Uemura scaling. A relation
between $T_c$ and $\rho(0)$ which combines the
effects of quasiparticle excitations at low temperatures
and of vortex fluctuations near critical temperature is proposed
and discussed in light of recent experiments.
\end{abstract}
\maketitle

High temperature superconductors appear most strikingly different from
those of standard BCS variety when underdoped. While the observed gap in
the single-particle density of states is large and appears to be
only increasing with  underdoping \cite{sutherland}, the $ab$-plane
superfluid density
at $T=0$, $\rho(0)$, at the same time is small and known to be
continuously decreasing
\cite{uemura}. The superconducting transition temperature $T_c$ appears
to scale approximately linearly
with $\rho(0)$, and this simple relation between the two fundamental quantities
has been ubiquitous in cuprates and widely known as the
Uemura scaling \cite{uemura}.
Recently, new measurements of the penetration depth 
in the strongly underdoped $YBa_2 Cu_3 O_{7-\delta}$ (YBCO)
suggested a significant 
deviation from the Uemura relation \cite{tami}, \cite{guikema}.
In this note we point out that the combination of three characteristics
of underdoped cuprates, namely quasi two-dimensionality,
a large superconducting
gap, and low $T_c$, irrespective of what the microscopic
mechanism behind the latter
two may be, severely constrains the form of $\rho(T)$.
In particular, we argue that even the early data of \cite{uemura} 
are difficult to explain quantitatively by the 
phase fluctuations alone, and that the quasiparticle contribution should make
the curve $T_c (\rho(0) )$ {\it convex}, as we believe better describes 
what is being experimentally observed \cite{tami}, \cite{guikema}, \cite{seaman}.

We begin from the early data on $T_c$ vs. $\rho(0)$
\cite{uemura}, \cite{singer}
by extracting a quasi-universal number that characterizes
this relationship \cite{singer}. The helicity modulus in a
two-dimensional (2D) superfluid layer
 may be expressed in units of Kelvin when written as
\begin{equation}
\rho(T) = \frac{\hbar ^2  d |\Psi_0 |^2}{k_B m^* },
\end{equation}
where $|\Psi_0 |^2 = m^* c^2 / 16\pi e^2 \lambda(T) ^2$ is the superfluid
density, $d$ a layer
thickness, and $\lambda(T) $ the in-plane penetration depth. Inserting all the
requisite constants one finds
\begin{equation}
\rho(T)= 6.2 K \frac{d/10\mbox{\AA}}{(\lambda(T) /\mu\rm{m})^2}. 
\end{equation}
In the case of layered quasi two-dimensional materials like cuprates there
is some ambiguity in defining the length $d$. Assuming it is the
minimal thickness along the c-axis so that the material still superconducts,
$d$ may be estimated from the measurements on superconductor-insulator
superlattices \cite{ivan} or from the crossing-point phenomenon
in magnetization measurements \cite{singer}. 
Using $10 \mbox{\AA} < d < 12 \mbox{\AA}$, with the lower boundary
corresponding to the
measurement of Goodrich et al. \cite{goodrich} on superlattices 
and the upper to the thickness of one c-axis unit cell in YBCO,
together with  $ T_c = (3.2 \pm 0.2 )(\mu\rm{m})^2 K / \lambda^2 (0)$
\cite{uemura}, \cite{singer}, one finds that
\begin{equation}
\frac{\rho(0)}{T_c} = Q,
\end{equation}
with the dimensionless amplitude 
 $1.82 <Q<2.48 $ in YBCO. Similarly, using $d=7.6 \mbox{\AA}$  \cite{singer}
and $ T_c = (3.0\pm 0.5) (\mu\rm{m})^2  K / \lambda^2 (0)$ \cite{singer},
\cite{tallon} one finds $1.37< Q < 1.88$ in $La_{2-x}
Sr_x Cu O_4$ (LSCO). Evidently, the amplitude 
$Q$ is not truly universal, but it is a pure number and most likely
around {\it two}.

First, we argue that assuming that phase fluctuations alone reduce 
the superfluid density in underdoped cuprates \cite{emery}, \cite{roddick}
leads to difficulties in accounting for the observed value of $Q$.
Consider an anisotropic quantum XY model describing thermal and quantum
fluctuations of the phase of the order parameter,
\begin{eqnarray}
S= -J\int_0 ^\beta d\tau \sum_{\vec{x},\hat{r},z }
\cos(\theta(\vec{x},z,\tau)-\theta(\vec{x}+\hat{r},z,\tau)) \\ \nonumber 
 - \alpha J  \int_0 ^\beta d\tau \sum_{\vec{x},z }
\cos(\theta(\vec{x},z, \tau)- 
\theta(\vec{x},z+1,\tau)) + \frac{1}{2 e_* ^2 } \\ \nonumber 
\int _0 ^\beta d\tau \sum_{\vec{x},\vec{y},z,z'}
\dot{ \theta} (\vec{x},z,\tau)
V^{-1} (|\vec{x}-\vec{y}|, z-z') \dot{\theta} (\vec{y},z', \tau), 
\end{eqnarray}
with $\vec{x}$ being a discrete two-dimensional (2D) vector labeling sites
of a quadratic lattice, $\dot{\theta}=\partial_\tau \theta$
and $\hat{r}=\hat{x}_1, \hat{x}_2$ a lattice unit vector.
The parameter $0<\alpha < 1$ measures the c-axis anisotropy.
The last term describes the quantum fluctuations induced by the
interactions and to be specific we will assume the Coulomb interaction 
$V\sim 1/\sqrt{|\vec{x}-\vec{y}|^2 + (z-z')^2 }$, although our arguments 
will be rather general.  For illustration,
we will also consider the short-range repulsion
$V=\delta_{\vec{x},\vec{y}}\delta_{z,z'}$.
In general, one can write the helicity modulus in the model (4) as
\begin{equation}
\frac{\rho}{\rho_0}= F( \frac{e_* ^2 }{\rho_0},
\frac{T}{\rho_0}, \alpha ) , 
\end{equation}
where $F$ is a dimensionless function of its three dimensionless arguments
 and $\rho_0 =\rho_0 (J,e_* ,\alpha) $ is the helicity modulus  at $T=0$.
(For $\alpha = e_* = 0$, for example, $\rho_0 =J$.)

   Let us consider first the 2D limit, $\alpha=0$. When $e_* =0$ as well,
the last term in (4) completely suppresses
quantum fluctuations, and the Eq. (4)
reduces to the standard (thermal) 2D XY model.  The critical temperature for
 $\alpha=e_* =0$ is
given by $T_c= \rho_0 / Q_{2D,XY}$, with $Q_{2D,XY}= 1.11$ \cite{gupta}.
The transition is in the Berezinskii-Kosterlitz-Thouless (BKT)
universality class, and consequently
at $T_c$ the superfluid density has a universal discontinuity, 
with $\rho(T_c^-)=(2/\pi) T_c$ \cite{nelson}.
At low temperatures the reduction of $\rho$ is
linear in $T$: $\rho (T) \approx \rho_0 - T/4$ \cite{gupta}.
 Consider now turning on a weak interaction $e_*$ while still keeping
 $\alpha=0$. This has two principal
 effects: 1) the reduction of $\rho_0 $ due to quantum
 fluctuations \cite{doniach}; 2) the introduction of a new (quantum)
 energy scale
 $\omega_q$ below which the {\it further} reduction of $\rho(T)$ with $T$ is
 suppressed. In principle, increasing the interaction  
 lifts the energies of the longitudinal (spin-wave)
 phase fluctuations, so $\Delta \rho (T) /\rho_0
 = (\rho_0 - \rho(T) )/\rho_0 $ should be
 a decreasing function of interactions (see Fig. 1).
 A weak Coulomb interaction $e_*$,
 nevertheless, is an irrelevant coupling in 2D even at $T=0$ \cite{herbut}
 so that the finite temperature transition remains
 in the BKT universality class and in particular, 
  $\rho(T_c^-)= (2/\pi) T_c$
 still holds for a weak $e_* \neq 0$. Defining the dimensionless amplitude 
 for $e_* \neq 0$ as $Q(\alpha=0,e_*)=\rho_0/ T_c$, the above discussion
 suggests then that
 \begin{equation}
 \frac{2}{\pi} < Q(\alpha=0,e_*) < Q_{2D, XY}, 
 \end{equation}
 which we propose here as a conjecture. 

\begin{figure}[t]
{\centering\resizebox*{80mm}{!}{\includegraphics{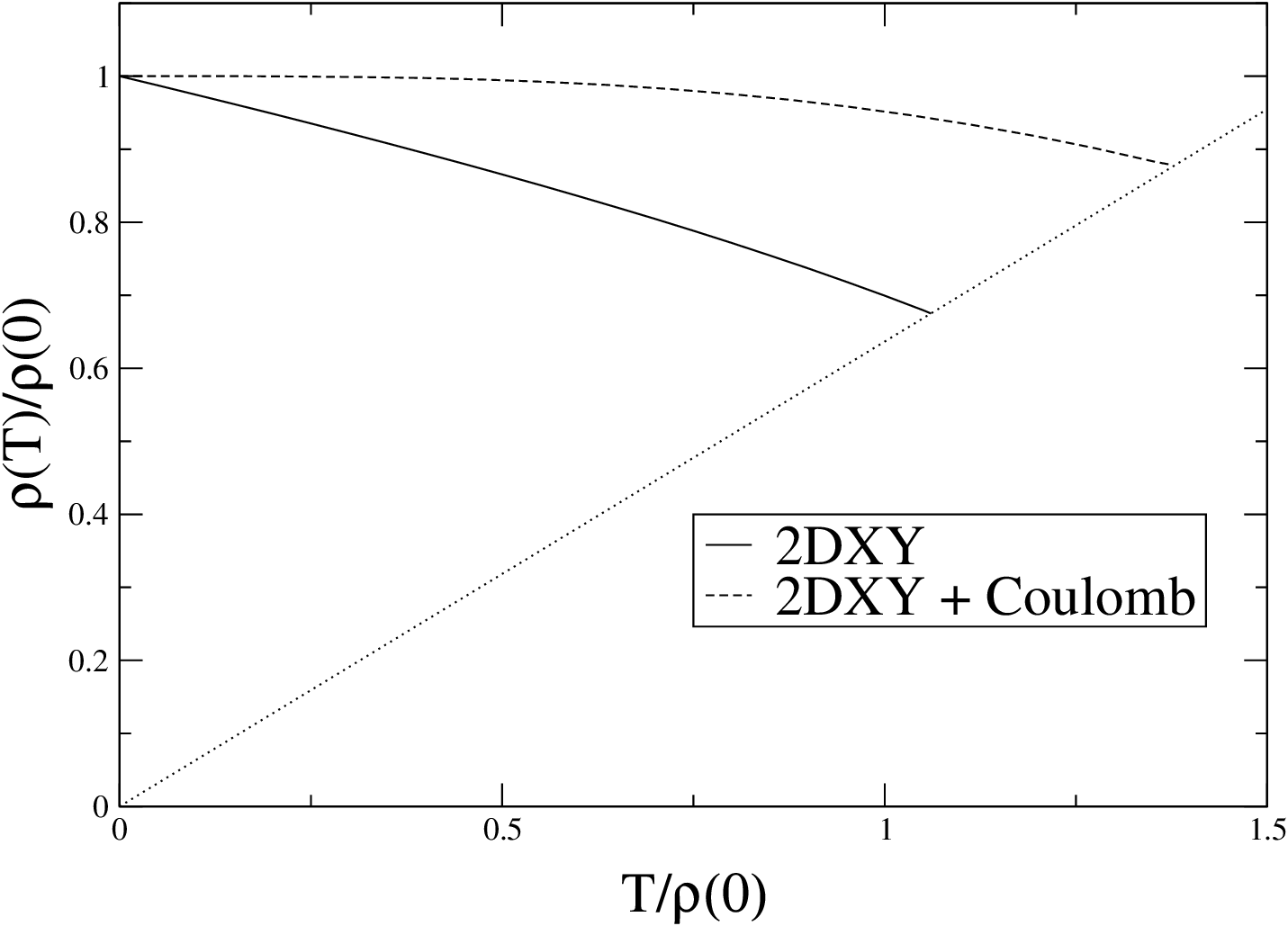}}\par}
\caption[]{Helicity modulus vs. temperature in the self-consistent
Gaussian approximation, for 2D XY model ($\alpha=e_* =0$), and
with Coulomb interaction ($\alpha=0$, $e_*^2=10$, and 
 $V(\vec{q})=1/|\vec{q}|$).
The straight dashed line denotes the universal BKT limit, 
 $\rho(T_c) = (2/\pi) T_c$, at which the helicity modulus discontinuously
 vanishes. Note that $T_c$ for fixed $\rho_0$
 {\it increases} with interaction.}
\label{RhovsT}
\end{figure}

 The lower bound in (6) is dictated by the BKT universality class 
 of the finite temperature transition in 2D. Our conjecture of the
 existence of the
 upper bound on $Q$ in two dimensions is supported by a self-consistent
calculation
 in which one approximates the first term in Eq. (4) (with $\alpha=0$)
 by an optimally chosen Gaussian term \cite{roddick}.
 The straightforward calculation then gives
 \begin{equation}
- ln \frac{\rho(T)}{J} = \frac{1}{4}\sqrt{\frac{e_*^2}{\rho(T)}}
\int \frac{d^2 \vec{q}}{(2\pi)^2}  F(\vec{q}) coth[\frac{\sqrt{e_* ^2
\rho(T)}} {T} F(\vec{q}) ],
\end{equation}
where $F^2 (\vec{q}) =  V(\vec{q}) \sum_{\hat{r}} sin^2 (\frac{\vec{q}
\cdot \hat{r}}{2})$. Identifying the approximate
transition temperature
as the point when classical vortices unbind and the self-consistent Gaussian 
approximation breaks down, $\rho(T_c) = (2/\pi) T_c$,
 at weak interaction Eq. (7) yields
\begin{eqnarray}
Q(\alpha=0, e_*) = \frac{2 e^{\frac{\pi}{8}}} {\pi}
[1- \frac{e_*}{4J^{\frac{1}{2}} } \int \frac{d^2 \vec{q}}{(2\pi)^2}  F(\vec{q}) \\ \nonumber
+ \frac{e_* ^2  e^\frac{\pi}{8}} {6 \pi J}
\int \frac{d^2 \vec{q}}{(2\pi)^2}  F^2 (\vec{q}) +O(e_* ^3) ] . 
\end{eqnarray}
Numerical results for $Q(\alpha=0,e_*)$ in our self-consistent Gaussian
approximation are shown in Fig. 2. In both the Coulomb and the
short-range case we find $Q$ to be a decreasing function of interaction.
Although the above calculation in principle includes only the
longitudinal (spin-wave)
phase fluctuations, it should be a reasonable indicator of the interaction
dependence of the amplitude $Q$. Recall that
the main contribution to $\Delta \rho(T)$ at temperatures
 $T\leq 0.8 T_c$ in the classical 2D XY model actually comes from  the 
spin-waves \cite{minnhagen}, and it is the increase of their energies with
interaction that should ultimately be responsible for the proposed 
behavior of $Q(\alpha=0,e_*)$.  

\begin{figure}[t]
{\centering\resizebox*{80mm}{!}{\includegraphics{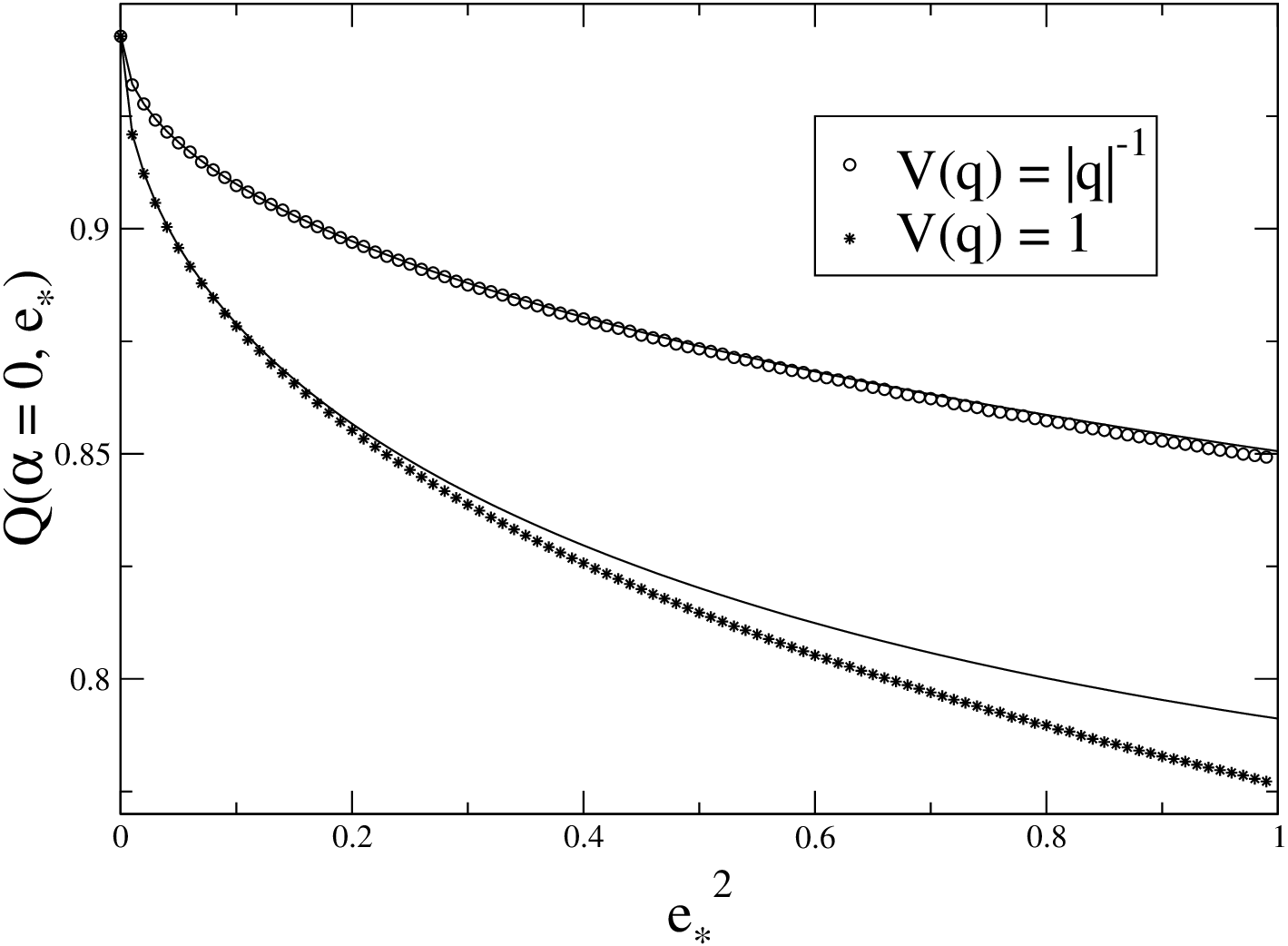}}\par}
\caption[]{The amplitude $Q(\alpha=0,e_*)=\rho_0 / T_c $ in the self-consistent
Gaussian approximation for the helicity modulus in 2D, Eq. (7),
with $T_c$ defined by
$\rho(T_c) = (2/\pi) T_c$, for a long-range and a short-range
interactions. The lines are the weak-interaction results in Eq. (8), and
$J=1$.}
\label{xymodels}
\end{figure}

At low temperatures $T\ll \omega_q $, from Eq. (7) one also
finds that for Coulomb interactions in 2D 
\begin{equation}
\frac{\rho(T)}{\rho_0} = 1- const. (\frac{T}{\omega_q})^5, 
\end{equation}
where $\omega_q = e_* ^{4/5} \rho_0 ^{3/5}$ is the characteristic
temperature scale for quantum fluctuations. For a short-range interaction one similarly
finds $1-(\rho(T)/\rho_0) \sim (T/\omega_q)^3$ in 2D, with $\omega_q =
e_* ^ {2/3} \rho_0 ^{2/3} $. Both expressions
show the anticipated reduction of
$\Delta \rho(T)/\rho_0 $ with interactions for fixed $\rho_0$.
Also, note that when $\rho_0 \rightarrow 0$, $\omega_q \gg \rho_0$ 
in general in 2D. This will have important consequences for the form of
$\rho(T)$, as we will discus shortly.

 The lesson from the above considerations is that interactions are expected 
 to {\it reduce}
 the amplitude $Q$ from its value in the 2D XY model.
 Introducing the third
 dimension by turning on $\alpha>0$ in (4) does the same. This is 
 evident in the Monte Carlo results \cite{carlson} as well as
 from the 3D value $Q_{3D, XY}= 0.45$ \cite{adler}, for example. The reason
 behind this trend is quite physical: coupling the layers only strengthens the
 phase coherence so that a higher temperature is required to destroy it.
 Both the interactions and the coupling between the layers
 should decrease $Q$ and we expect that 
 \begin{equation}
 Q (\alpha, e_*) \leq Q_{2D, XY}=1.11
 \end{equation}
for the family of quantum XY models defined by Eq. (4).
Introducing longer-ranged couplings than just
between the nearest
neighbors \cite{carbotte}, as long as there is no frustration in the
problem \cite{remark}, would also increase the tendency to order
and consequently should reduce $Q$ further.
The maximum value of $Q_{2D,XY}$ then corresponds to the weakest
(unfrustrated) superfluid order on the quadratic lattice \cite{remark1}.

 The conjectured maximum value of the ratio $Q$ in the family of quantum 
 XY models (4) appears to be significantly below
the one characterizing the standard experimental
Uemura plot. To bring the two
closer one needs to assume a considerably smaller
`thickness' $d$: choosing the average spacing between the adjacent
layers in YBCO, $d=6 \mbox{\AA} $, for example, would lead to 
 $Q> 1.09$ from experiment, and possibly falling just slightly below  our 
upper bound. A more natural interpretation of the above discrepancy is that
an additional mechanism for the reduction of superfluid density is
required. Of course, one is readily available in cuprates: nodal
quasiparticles.

As well known \cite{won,lee},
 with the phase fluctuations neglected, at low temperatures
depletion of the helicity modulus in a 2D d-wave
superconductor is entirely due to low-energy quasiparticles near the four
nodes of the order parameter,
\begin{equation}
\rho= \rho(0) - y T + O(\frac{T^2}{\Delta}).
\end{equation}
The number $y= (2 ln(2)/\pi) z^2 (v_F/v_\Delta)$, $\Delta$ is
the maximum of the superconducting gap, $v_F$ and $v_\Delta$ are the two
velocities characterizing the low-energy quasiparticle
dispersion, and $z$ is the
charge renormalization factor. Assuming $\Delta\gg T_c $ in underdoped
cuprates makes the correction to linear dependence in Eq. (7) 
negligible for $T<T_c$ and we drop it hereafter.

\begin{figure}[t]
{\centering\resizebox*{80mm}{!}{\includegraphics{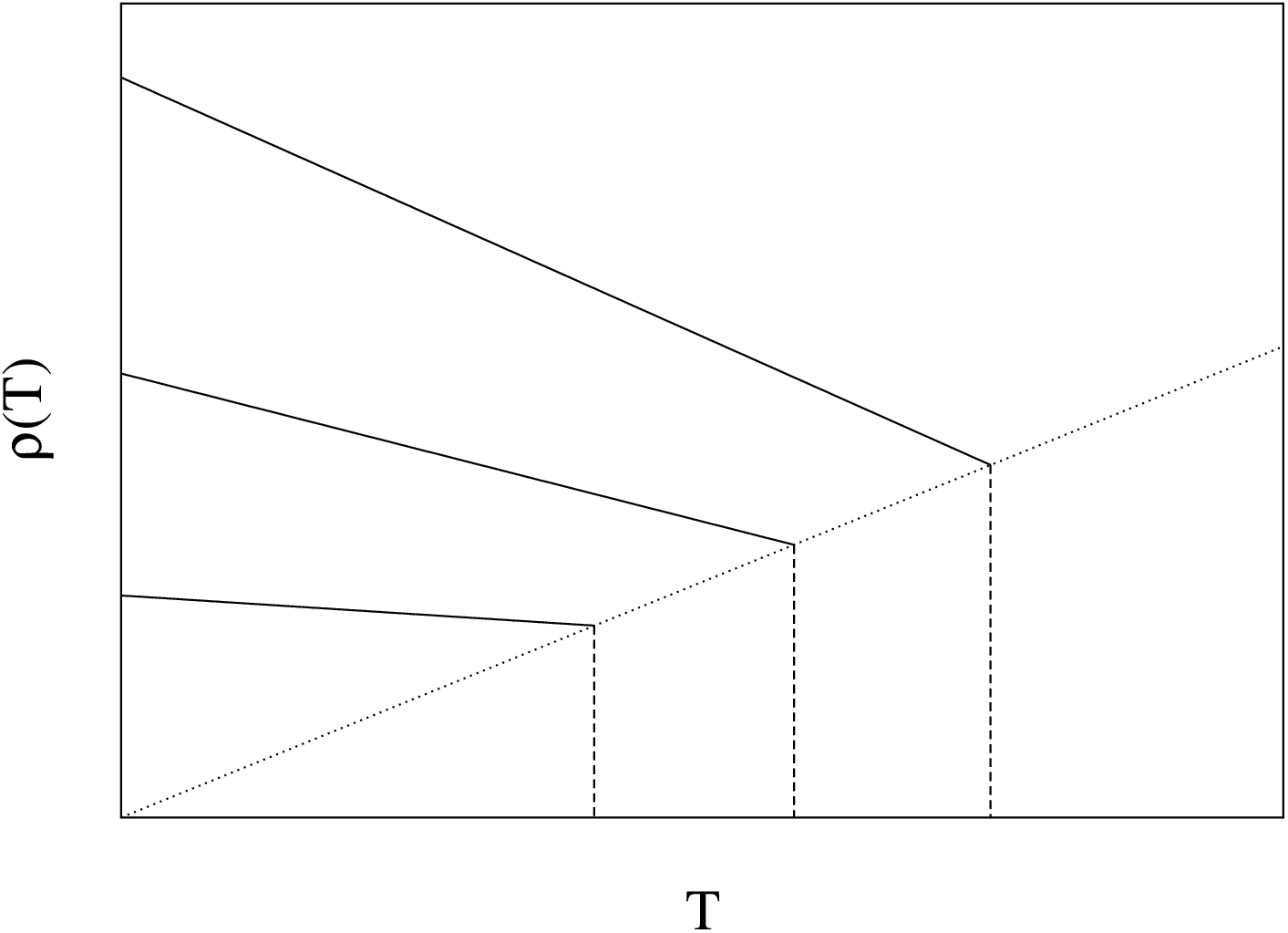}}\par}
\caption[]{Schematic behavior of the
superfluid density vs. temperature in a low-$T_c$, large
gap, quasi-2D d-wave superconductor. Doping is assumed to be decreasing
from top to  bottom. Note that both $T_c$ and the slope
should decrease with underdoping (see the text).
Dashed line again represents the universal BKT limit.}
\label{dwavesf}
\end{figure}

Very underdoped cuprates are characterized by a large gap $\Delta \gg T_c $,
small transition temperature $T_c$, and strong c-axis anisotropy. 
 We may then include both the quasiparticle and phase
 fluctuation contributions to the reduction  of the superfluid density
 by reasoning as follows: 1) Assume that the $T=0$ superconducting
 transition in underdoped regime is continuous. General scaling
 arguments imply then that {\it near} the quantum critical point
  $\rho(0)\sim T_c ^{(D+z-2)/z}$, which for
 $D=2$ (2D) reduces to $\rho(0)\sim T_c$ independently of the nature
 of the putative quantum critical point. It then follows that as $T_c
 \rightarrow 0$, the quantum temperature scale in Eq. (9)
 $\omega_q \gg T_c$ and the longitudinal 
 phase fluctuations are strongly suppressed for {\it all} $T<T_c$. 
 2) With phase fluctuations being essentially negligible,
 quasiparticles alone reduce $\rho(T)$ according to Eq. (11), with only the
 linear term being important for $T< T_c \ll \Delta$. 3) Finally,
 the superfluid density in 2D still can not fall below the universal
 BKT value, since for $\rho(T)< (2/\pi) T$ the superfluid order would become 
 unstable to unbinding of classical vortices.
 In sum, the transition temperature in a quasi-2D 
 d-wave superconductor with $\Delta \gg T_c$ is approximately
 determined by
 \begin{equation}
 \rho(0)-y T_c = \frac{2T_c }{\pi}, 
 \end{equation}
 where $\rho(0)$ is the {\it physical} $T=0$ helicity modulus, already renormalized
 by quantum fluctuations. $\rho(T)$ at various dopings
 acquires then an approximate simple form depicted schematically 
 in Fig. 3. Virtual (bound) vortex fluctuations
 should additionally reduce $\rho(T)$ near $T_c$, and thus tend to
 slightly lower the transition temperature. 
 With $0< \alpha \ll 1$, on the other hand, some rounding of the
 universal BKT jump close $T_c$ is expected, as well as
 a slight increase of $T_c$. These fine effects are being omitted from
 Fig. 3 for simplicity but should be expected in experiments.

\begin{figure}[t]
{\centering\resizebox*{80mm}{!}{\includegraphics{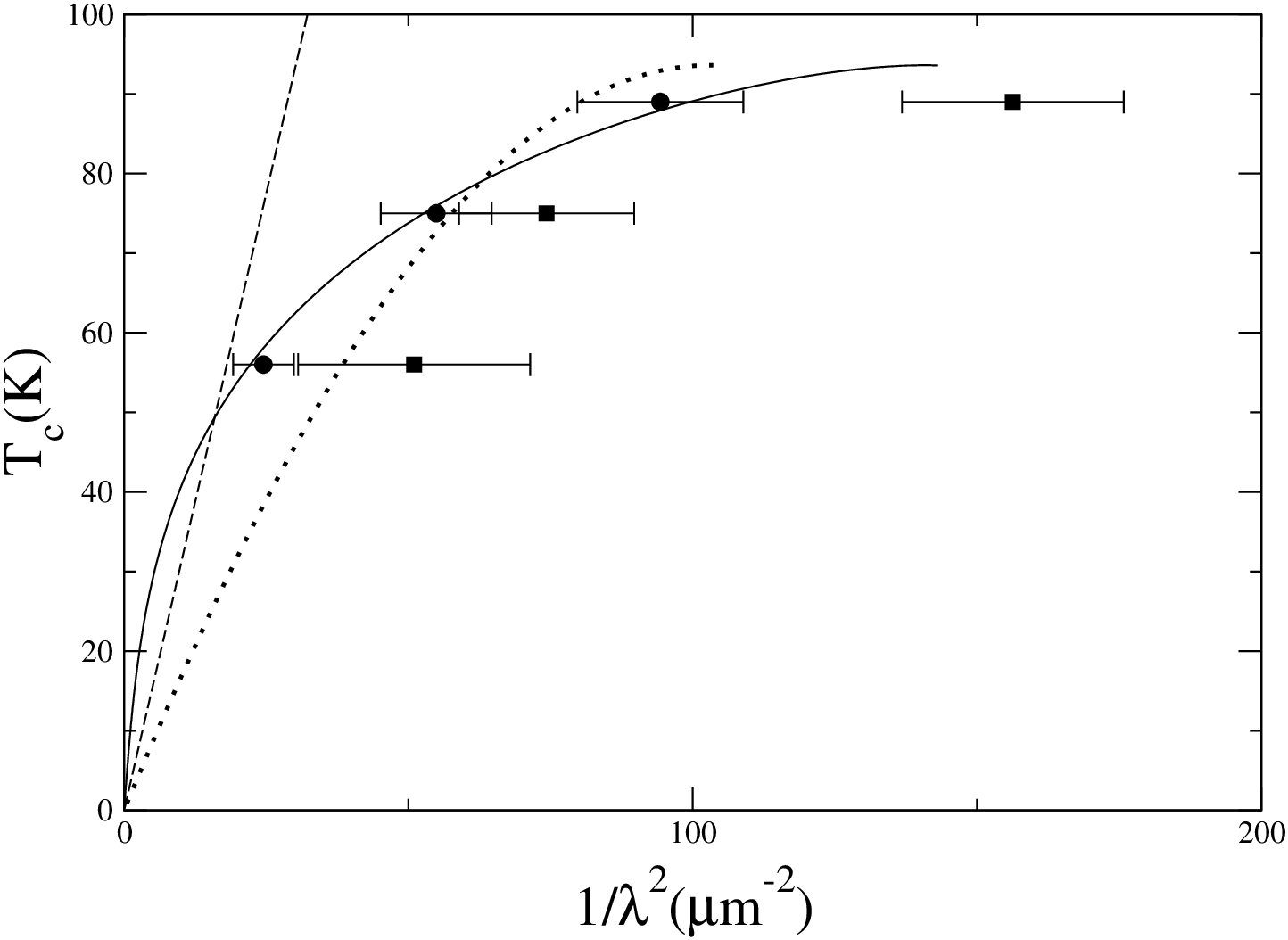}}\par}
\caption[]{Data reported in \cite{tami}.  The lines are fits to the data as
discussed in the text.  The convexity of the curves is due to the doping
dependence of the quantity $z^2(v_F/v_{\Delta})$.}
\label{uemurafit}
\end{figure}

The addition of the quasiparticle contribution  $y$ in  Eq. (12) makes the
agreement with old
Uemura data immediately better: taking $y\approx 1$ near optimal
doping \cite{bonn} brings 
the sum in $Q=y+(2/\pi)$ closer to two. More importantly, 
Eq. (12) implies a systematic deviation of  
$T_c (\rho(0))$ from the simple linear relation due to the anticipated 
decrease of the slope $y$ with underdoping. Even if one assumes
the charge renormalization factor $z$ to approximately stay constant
with doping, the decrease of $v_\Delta$ \cite{sutherland} 
together with the approximate constancy of $v_F$ \cite{shen}
makes the slope $y$ inevitably an increasing
function of doping, and therefore of $\rho(0)$. The slope of the
$T_c (\rho(0))$ curve should therefore be a decreasing function of
$\rho(0)$, i. e. the curve should be {\it convex} upward. In fact  we 
expect that the dependence of $y$ on doping should be stronger
than what would follow from the observed increase of $v_\Delta$ alone.
This would be in agreement  with the general theory of the
fluctuating d-wave superconductors \cite{igor}
and  the gauge theory of t-J model \cite{lee},
which both predict the charge renormalization factor $z\propto T_c$ at
small dopings. 

Figure 4 shows the data taken from Pereg-Barnea {\it et al.} \cite{tami} where
circles (squares) are from $a(b)$-axis measurements.  To avoid contributions
arising from chains in the $b$-direction, we consider only the
$a$-direction, though both data sets clearly show a departure from linear
behavior. In fitting the data we used a straight line fit to the doping
dependence of the superconducting gap $\Delta_0$ from Sutherland {\it et al.}
(their Figure 6) \cite{sutherland} and converted to $v_{\Delta}$ using
$\Delta_0 = \hbar k_F v_{\Delta}/2$ with $k_F \approx 0.7 \mbox{\AA}^{-1}$.
We have taken $v_F = 2.5\times 10^7$ cm/s.  The dotted line
represents then the Eq. 12 with $d=10\mbox{\AA}$
and $z=1$ and no fitting parameters.
A somewhat better agreement is obtained by assuming $z = T_c/C$ with $C=79$K,
which we show as the full line, as an illustration. The reader  is 
cautioned, however, that such a strong
dependence of the slope $y$ on $T_c$ has not yet been directly
observed. The dashed line is the linear Uemura relation with
$Q=2$, demonstrating that the data departs significantly from this
behavior.

The curvature in the $T_c(\rho(0))$ plot was also noted and discussed in a
recent paper by Tallon et al. \cite{tallon} where it was attributed
to the opening of an additional gap in the density of states {\it competing}
with superconductivity. Our explanation is fundamentally different in nature
and in particular, the curvature is argued to be a generic feature of
two-dimensional d-wave superconductor with the increasing gap and the
decreasing $T_c$. The
microscopic mechanism behind the latter two properties of
underdoped cuprates is left unaddressed here 
and will be a subject of a separate publication \cite{igorunpub}. 

In sum, we discussed the reduction of superfluid density in a quasi-2D
d-wave superconductors with low $T_c$ and large superconducting gap.
We argued that the Coulomb interaction under these conditions essentially
eliminates the longitudinal phase fluctuations but that the transition
at $T_c$ should still be governed by the universal vortex physics
and fall into the BKT universality class. A relation
between $\rho(0)$ and $T_c$ which takes into account both the
quasiparticle contribution and the universal vortex
fluctuations near $T_c$ is proposed
to explain the deviation from the linear Uemura scaling observed 
in recent experiments on underdoped YBCO. 

We thank D. Bonn, I. Bo\v zovi\' c, D. Broun,
 and P. Turner for useful discussions.
This work was supported by the NSERC of Canada and the Research Corporation.


\begin{thebibliography}{99}
\bibitem{remark} We are assuming that the additional terms
$\sim J_n \cos(n(\theta(\vec{x},z)-\theta(\vec{y},z')) )$
with $n \geq 2$ in Eq. (4), which effectively introduce such frustration
(see ref. 17), are negligible.
\bibitem{remark1} We did not include the `Berry phase' term
$i n_0 \int d\tau \sum_{\vec{x},z} \dot{\theta}(\vec{x},z,\tau)$
in Eq. (4). At low temperatures such term vanishes for the 
spin-wave phase fluctuations, and it is also irrelevant at the BKT
transition at $T_c$. We therefore do not expect it to affect our arguments 
leading to the inequality (10).
\end{thebibliography}
\end{document}